\documentclass{aipproc}

\layoutstyle{8x11double}

\SetInternalRegister\hbadness{8000} 

\newcommand\doingARLO[2][]{%
  \ifx\mmref\undefined #1\else #2\fi
}

\begin{document}

\title[X--ray spectroscopy of GRBs]
{X--ray spectroscopy of $\gamma$-ray bursts: the path to the progenitor}


\author{Davide Lazzati}{
address={Institute of Astronomy, Madingley Road CB3 0HA Cambridge, U.K.}
}

\iftrue
\author{Rosalba Perna}{
  address={Harvard Smithsonian Center for Astrophysics, 60 Garden street,
Cambridge MA, 02138}
}

\author{Gabriele Ghisellini}{
  address={Osservatorio Astronomico di Brera, via Bianchi 46, 23807 
Merate (LC), Italy}
}
\fi

\copyrightyear  {2001}

\begin{abstract}
Despite great observational and theoretical effort, the
burst progenitor is still a mysterious object. It is generally
accepted that one of the best ways to unveil its nature is the study
of the properties of the close environment in which the explosion
takes place. We discuss the potentiality and feasibility of time
resolved X--ray spectroscopy, focusing on the prompt $\gamma$-ray phase. 
We show that the study of absorption features (or continuum
absorption) can reveal the radial structure of the close environment, 
unaccessible with different techniques. We discuss the detection of 
absorption  in the prompt and afterglow spectra of several
bursts, showing how these are consistent with gamma-ray bursts taking
place in dense regions. In particular, we show that the radius and
density of the surrounding cloud can be measured through the evolution
of the column density in the prompt burst phase. The derived cloud
properties are similar to those of the star forming cocoons and
globules within molecular clouds. We conclude that the burst are
likely associated with the final evolutionary stages of massive stars.
\end{abstract}

\date{\today}

\maketitle

\section{Introduction}
It is widely believed that a good way to understand which is the 
progenitor of GRBs is by analyzing the properties of the interstellar 
medium that surrounds the explosion. This is because the fireball early
self--similar evolution erases all the traces of its initial condition 
and hence any pre and during--explosion signature.

The three main classes of burst progenitor can be, in principle, 
easily distinguished by mean of the properties of their environment.
If the burst are due to the merger event of binary coalescing systems 
of neutron stars \citep{eic89} they are expected to 
take place in a uniform low density ($n\sim0.1-10$~cm$^{-3}$) intergalactic 
medium. This is due to the fact that the binary system has a long life 
($\sim 10^9$~y) and a high proper motion ($v\sim100-1000$~km/s) and can 
travel out of the original birth place before the merging event (but see 
Perna et al., this volume)

If the explosion of a GRB is coincident with the explosion of a massive
rotating star (hypernov\ae~or collapsar, Woosley, this volume), it has to be 
surrounded primarily by the pre--explosion stellar wind. This wind will then 
impact on the molecular cloud in which the star was born, with a shock 
contact (terminal) discontinuity \citep{ram01}.

Finally, the bursts may be associated to supernova explosions but
with some delay (see the supranova model \citep{vie98}). In this case
the burst should explode in an evacuated cavity, surrounded by a supernova 
shell and eventually by a molecular cloud medium.

These three radial profiles of the ambient media surrounding GRBs are 
sketched in the left panel of Fig. 1, where the solid line represents 
compact mergers and the dashed and dotted lines represent hypernovae 
and supranovae, respectively.

In principle the density profile can be traced by modelling the afterglow 
light curves and spectra. One must however be aware that
most afterglow data are taken between half a day and several months
after the burst explosion. As it is shown in the right panel 
of Fig. 1, in this period of time the fireball, no matter the progenitor 
model, runs through a uniform medium, the only difference being the 
normalization (probably the most uncertain of all the model parameters).
The morphological difference of the left panel is then impossible to 
reconstruct with present day measurements.

There are several alternatives in order to measure the close environment 
density and structure. One is to be fast. In principle, if one can have 
(as we will have in the Swift era) detailed early time light curves, the 
whole radial density structure can be measured. However, it is likely that 
the emission mechanism of the afterglow gets more complicated as we approach 
the explosion site: reverse shock emission, late injection of energy from 
the inner engine and the superposition with the radiation
from internal shocks will probably make 
the modeling of early time afterglows a delicate issue.

\begin{figure}
  \includegraphics[width=\textwidth]{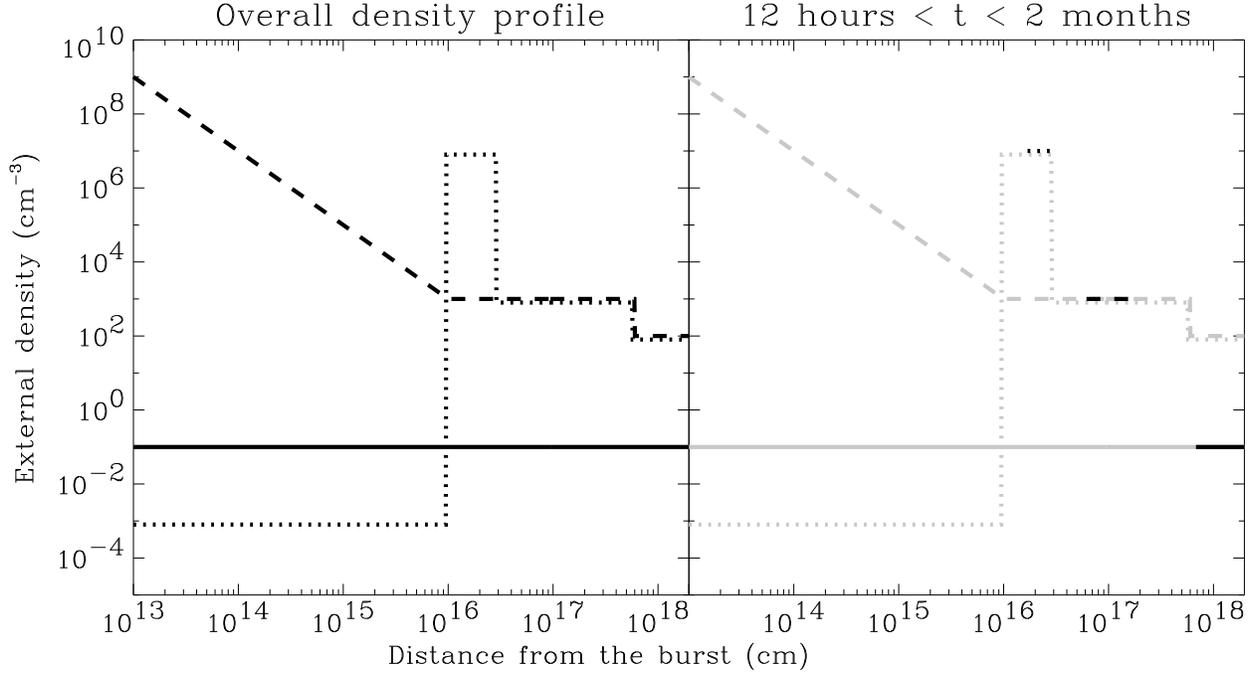}
  \caption{Radial density profiles for different GRB progenitors.
the solid line represent 
compact mergers and the dashed and dotted lines represent hypernovae 
and supranovae, respectively. The left panel shows the pre--explosion 
setup, while in the right panel only the range of radii that the fireball 
travels between an observer time $t=12$~hours and $t=2$~months is highlighted.}
\end{figure}

\begin{figure}
  \includegraphics[width=0.48\textwidth]{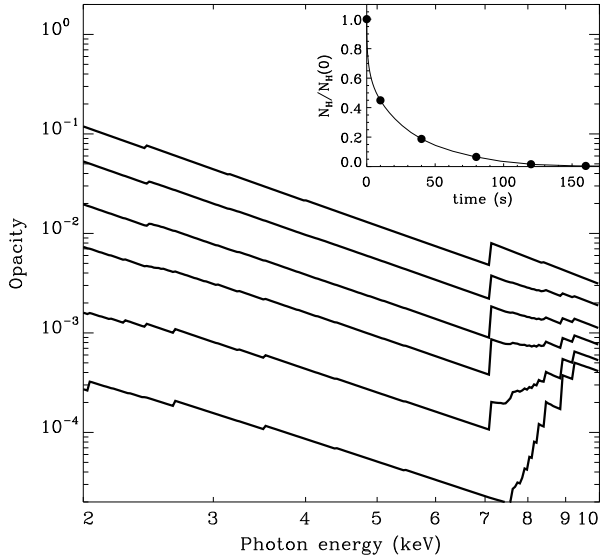}
  \caption{Opacity in the range [2-10]~keV for a cloud with solar
metallicity, $R=3\times10^{18}$~cm and initial column density
$N_H(0)=3\times10^{21}$~cm$^{-2}$. In the main panel, from top to
bottom, we plot the absorption at times $t=0$, 10, 40, 80, 120 and 160
seconds. In the inset, the column density is shown as a function of
time. Filled dots mark the column densities corresponding to the
spectra plotted in the main panel.}
\end{figure}

An alternative is to look for echoes, i.e. photons that, initially emitted
at large angles with respect to the line of sight, are scattered in the 
direction of the observer. Dust \citep{esi00}, Compton 
\citep{mad00} and iron line \citep{laz99} echoes have been 
proposed. Only iron lines have been securely observed, to date 
\citep{pir00}. The modelling of echoes presents two difficulties: 
first, if GRB fireballs are highly collimated, there is little room for 
echoes. Secondly, it is difficult to disentangle photons scattered by a 
large angle at small distance from the burst site from photons scattered 
at small angles at a larger distance from the progenitor.

We here propose and analyze a method, based on prompt time--resolved 
X--ray spectroscopy of the burst photons, which is unbiased and rely on 
very well known physics. The propagation of the photons
in the ambient medium will in fact imprint absorption features on
the soft X--ray spectra. These features will become less deep as the ionization 
front expands, allowing us to measure the density and the radial profile 
of the surrounding material.

\section{Column density evolution}

There are in principle two ways of exploiting time resolved X--ray 
spectroscopy. In case of very good quality data, one can follow the 
opacity vs. time of a single, well isolated feature. An example is the
iron $K_\alpha$ photoionization edge \citep{laz01a,laz01b} at 
7.1 keV. This has the advantage of being completely model independent.
On the other hand, the opacity of the iron edge for solar metallicity 
material is very small, and a Thomson thick cloud is necessary in order 
to observe a $\tau_{\rm Fe}>1$ feature.

\begin{figure}
  \includegraphics[width=0.48\textwidth]{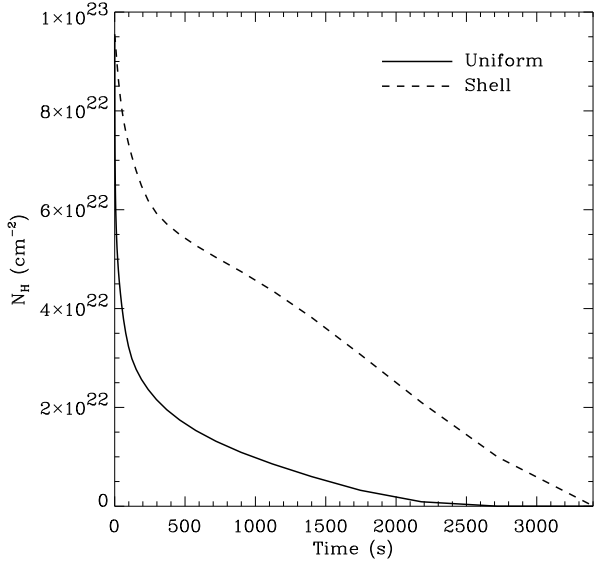}
  \caption{Evolution of the column density with time for a uniform
(solid line) and shell (dashed line) environments. In both cases the
initial column density is $N_H(0)=10^{23}$~cm$^{-2}$ and the outer
radius is $R=1$~pc.}
\end{figure}

At lower energies ([0.1-2]~keV), the spectra are more crowded and it 
is very difficult to follow a single transition. However, the opacity 
is much larger, and even a $\tau_T<0.01$ cloud can yield a very easily 
measurable signal. Usually the quantity of absorbing material is 
parametrized through the quantity $N_H$, i.e. the column of solar 
metallicity cold material that would absorb the same quantity of 
soft X--ray photons.

In the case of material surrounding GRBs, the assumption of solar 
metallicity can be reasonable, but the measured $N_H$ is much different 
from the real column density due to the progressive ionization the medium 
undergoes as the burst photons propagate through it. In order to 
estimate the amount of absorbing material as a function of time
(which will show as $N_H$ in the spectra) 
we run many photoionization simulations
(see \citep{lazper} for more details). Fig. 2 shows the result of 
one of these simulations, in terms of the frequency--resolved time dependent 
opacity (main panel) and of the measured column density (inset). The advantage 
of the method is that different radial density profiles will 
give different time evolutions of the $N_H$ evaporation. 
For example, Fig. 3 shows the case of a uniform cloud and of a shell with 
the same initial column density. 
The shell material is more difficult to
photoionize and the column density evolution is then slower.

\subsection{Application to GRB data}

Even though with limited spectral resolution and statistical quality, 
some time resolved column density measurement have been performed 
with real data. We show in Fig. 4 and Fig. 5 the cases of GRB~980329
\citep{fro00}, observed with {\it Beppo}SAX and GRB~780506
\citep{con98}. In both cases a fairly dense and compact
region is derived as a best fit to the data. The radial profile could not
be measured since in one case (GRB~980329) the error bars of the measurement
were too large and in the other (GRB~780506) only one positive detection was 
made.

\begin{figure}
  \includegraphics[width=0.48\textwidth]{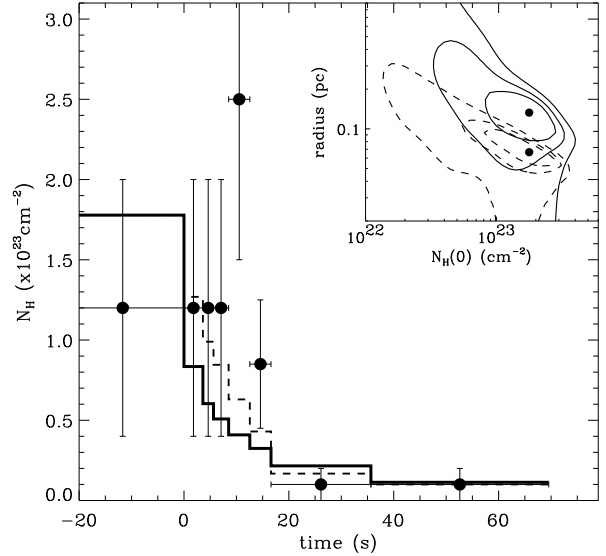}
  \caption{Evolution of the column density measured in the prompt X--ray
spectrum of GRB~980329 \citep{fro00}. A uniform and shell
geometry for the absorbing column have been tested. The best fit
models, integrated over the same time intervals of the data, are shown
with a solid and a dashed line, for the uniform and shell geometries,
respectively. The inset shows the $1\sigma$, 90 per cent and 99 per
cent confidence contours for the two fitted parameters. Again, solid
contours refer to uniform density and dashed contours to a shell
geometry.}
\end{figure}

\begin{figure}
  \includegraphics[width=0.48\textwidth]{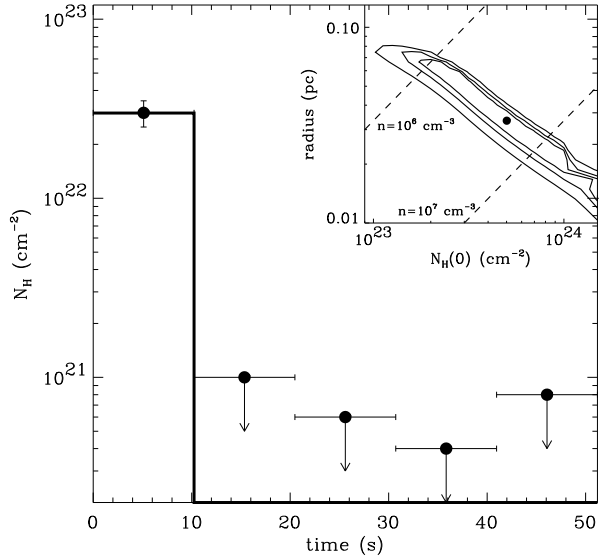}
  \caption{Same as Fig.~4 but for GRB~780506
\citep{con98}. The solid line shows the best fit model with $N_H(0)=
2\times10^{23}$~cm$^{-2}$ and $R=2\times10^{17}$~cm integrated over
the observed time intervals. The inset shows the $1\sigma$, 90 per
cent and 99 per cent contour levels for the two parameters and, with
dashed lines, isodensity contours for $n=10^6$ and $10^7$~cm$^{-3}$.}
\end{figure}

\section{discussion}

We have shown that time resolved X--ray spectroscopy of the early phases 
of GRB emission can give us informations on the density and radial 
structure of the surrounding material.
Given the capabilities of present days instrumentations,
this can be effectively done in case of fairly dense and compact regions.
If we impose that the column density must not be negligible after 1 second 
of observation and that it must decrease by a factor of two after 100 
seconds of GRB emission, we find that, for a uniform absorbing cloud, 
positive detections of $N_H$ variations should be performed if a GRB is 
surrounded by a cloud with size and column density marked with the gray 
shading in Fig. 6.

We compared these cloud properties with typical properties of molecular 
clouds and their overdense regions in our Galaxy. We find that if GRBs 
take place in random locations inside molecular clouds, their X--ray early
spectra should show no sign of photoionization absorption,
since all the material is ionized on a time scale of less than one second.
On the other hand, massive stars are thought to be born inside overdense 
and compact regions within molecular clouds. If GRBs are associated with 
these regions, evolution of the X--ray absorbing column should be detectable.
In particular, the variable absorption observed in the spectra of 
GRB~980329 and GRB~780506 can be explained if they were located in 
regions with properties close to those of Bok globules.

\begin{figure}
  \includegraphics[width=0.48\textwidth]{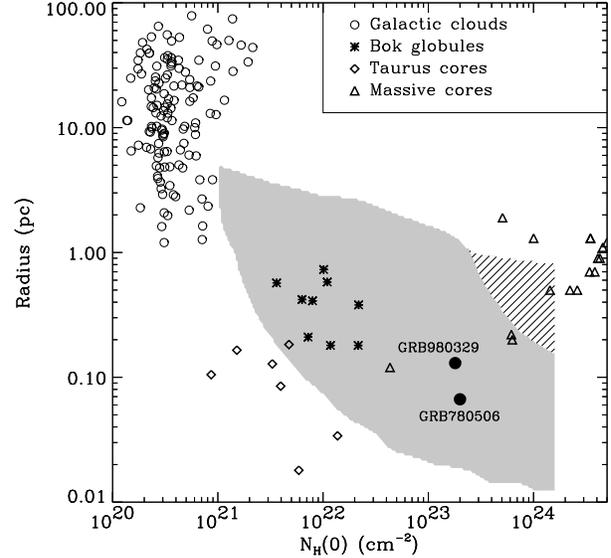}
  \caption{Comparison between the radius and column densities inferred 
 and the properties of Galactic molecular clouds and
some hierarchical structures embedded in them. The gray and line
shaded areas are where variable column is observable with GRB measurements. 
Circles show radii and average column densities
of a sample of Galactic molecular clouds. 
Asterisks refer to Bok globules,
diamonds refer to dense cores in the Taurus molecular cloud 
and triangles refer to massive cloud cores. 
The two filled dots are the best fit values to the $N_H$
measurements of GRB~780506 and GRB~980329.}
\end{figure}


\doingARLO[\bibliographystyle{aipproc}]
          {\ifthenelse{\equal{\AIPcitestyleselect}{num}}
             {\bibliographystyle{arlonum}}
             {\bibliographystyle{arlobib}}
          }
\bibliography{nh_WH}

\end{document}